# The emergence of unshared consensus decisions in bottlenose dolphins[1]

David Lusseau[†] and Larissa Conradt[‡]

[†]University of Aberdeen, Institute of Biological and Environmental Sciences, Aberdeen, UK. E-mail: d.lusseau@abdn.ac.uk.

[‡]University of Sussex, Department of Biology and Environmental Science, Brighton, UK. E-mail: L.Conradt@sussex.ac.uk.

**ABSTRACT**

Unshared consensus decision-making processes, in which one or a small number of individuals make the decision for the rest of a group, are rarely documented. However, this mechanism can be beneficial for all group members when one individual has greater knowledge about the benefits of the decision than other group members. Such decisions are reached during certain activity shifts within the population of bottlenose dolphins residing in Doubtful Sound, New Zealand. Behavioral signals are performed by one individual and seem to precipitate shifts in the behavior of the entire group: side flops are performed by males and initiate traveling bouts while upside-down lobtails are performed by females and terminate traveling bouts. However, these signals are not observed at all activity shifts. We find that while side flops were performed by males that have greater knowledge than other male group members, this was not the case for females performing upside-down lobtails. The reason for this could have been that a generally high knowledge about the optimal timing of travel terminations rendered it less important which individual female made the decision.

**Keywords**

behavioral ecology, decision-making process, bottlenose dolphin, group living

## INTRODUCTION

Social animals need conspecifics to face a multitude of challenges, from predation pressure (Heard 1992; Noe and Bshary 1997) and food acquisition (Baird and Dill 1996; Fritz and de Garine Wichatitsky 1996; Blundell et al. 2002; Lusseau et al. 2004) to intra-specific competition (Yamagiwa et al. 2003; Jakob 2004) and mating competition (Connor et al. 2001; Boyko et al. 2004). Groups of social animals have to reach consensus in order to synchronize the activity of their members and maintain the cohesion of the group (Ruckstuhl 1999; Conradt and Roper 2003; Couzin et al. 2005). Shared consensus decisions, in which the group does what the majority of its members want, seem to be prevalent in a wide variety of taxa from insects to primates (Conradt and Roper 2005). Unshared consensus, previously referred to as despotic decisions, in which one individual makes the decision for the rest of the group, are rarer but can be beneficial in certain circumstances (Conradt and Roper 2003).

Synchronizing activities in a group is the result of a cost-benefit analysis performed by each individual. Different classes of individuals have different metabolic requirements and therefore this trade-off assessment often raises conflicts of interest between group members (Conradt and Roper 2005). For example, it can be costly for an individual to start traveling if it has not finished foraging. This differential in requirements leads to sexual segregation in the association patterns of at least some ungulates when there is sexual dimorphism in body size (Conradt and Roper 2000; Ruckstuhl and Neuhaus 2002). However, if the majority of the group wants to start traveling, an individual will lose the benefits of belonging to the group if it stays behind and the benefits of group membership can often outweigh the costs of not fully meeting the individual's metabolic requirements (Conradt and Roper 2003). In these situations, unshared decisions are only likely to evolve if the individual making the decision has a greater knowledge about the benefits of shifting activities than all other members together (List 2004). In these instances, other group members need to know that following the decision of this temporary 'leader' will incur them a smaller fitness cost than their own decision would (Templeton and Giraldeau 1996; Conradt and Roper 2003). The leader may have a better understanding of the cost-benefit of the current patch occupied by the group or that individual may have more knowledge about other patches and can therefore assess more accurately the trade-offs sustained by leaving the current patch (Templeton and Giraldeau 1996). These findings are derived from recent modelling work which provides a framework of testable hypotheses about when we should expect unshared consensus decisions to emerge (Conradt and Roper 2005). One of the empirical difficulties to test these models is in finding a way to estimate the relative position of individuals in relation to others in the population in order to account for potential discrepancies in knowledge. Social network analyses offer statistics to quantify these positions (Lusseau et al. 2008).

---

[1] This contribution is part of the special issue "Social Networks: new perspectives" of *Behavioral Ecology and Sociobiology* (guest editors: J Krause, D Lusseau and R James )

There are early indications from observational studies in Doubtful Sound, New Zealand, that the bottlenose dolphin (*Tursiops* sp.) population residing in this fiord is using unshared consensus decision-making in some instances (Lusseau 2006; Lusseau 2007a). Two behavioral cues are significantly more likely to be observed only during activity shifts: side flops (SF) as the group (school hereafter to avoid confusion with social group/unit terminology) initiates traveling bouts and upside-down lobtails (ULT) when the school terminates traveling bouts (Lusseau 2006). We do not know if these events are visual, acoustic, or multi-sensorial cues. These behavioral events are always only performed by one individual in the school and *ad libitum* identification of the behavior performers showed that side floppers are males and upside-down lobtailers are females in almost all cases (Lusseau 2007a). In addition, these individuals share common positions in the social network of the population (Lusseau 2007a). They tend to be social brokers, individuals who spend time with individuals from different social units in the population, hence individuals who have an understanding of the past and current activities of a greater number of individuals and social units than most others in the social network. This means that these individuals are likely to have a greater understanding of both foraging patches recently visited, and depleted, by others and the status of potential competitors. This is important for this population because we know that these dolphins rely almost exclusively on spatially-fixed, reef-associated, prey items (Lusseau and Wing 2006).

However, these behavioral cues are not performed during all activity shifts (Lusseau 2006). Here, we firstly hypothesize that dolphins are more likely to reach unshared consensus decisions about travel (in particular, the start of travel) if the discrepancy in knowledge between one individual and others in the school is high. Secondly, we test our predictions by investigating whether apparent unshared consensus decision-making increased with discrepancy in knowledge.

## METHODS

**Field techniques**

Behavioral data were collected in Doubtful Sound, New Zealand (45°30' S, 167°00' E) between April 2000 and April 2002. Systematic surveys of the fiord were conducted to look for dolphin schools (Lusseau et al. 2003). Once a school was detected the identity of individuals in the school was determined using photo-identification (Würsig and Würsig 1977). A code of conduct was established for the observing vessel to minimize its effects on the focal schools (Schneider 1999). Studies showed that the behavior of the focal schools was not affected by the presence of the observing vessel (Lusseau 2003; Lusseau 2007a). Side flops (SF) and upside-down lobtails (ULT) are rare events (0.72 sf/hour and 0.96 ult/hour of focal follows (Lusseau 2006)); we therefore recorded the occurrence of side flops and upside-down lobtails in an *ad libitum* fashion while following the school (Altmann 1974; Mann 2000). Side flops were defined as jumps during which a dolphin cleared its entire body out of the water and landed on its side. Upside-down lobtails were defined as situations when a dolphin was upside-down stationary at the surface, belly pointing upwards, and forcefully slapped the water surface with its tail. Observations ended when the weather deteriorated, the focal school was lost, or the day ended, therefore the end of an observation period was not dependent on the behavior of the focal school.

The gender of photo-identified individuals was assessed by direct observation of the genital slit using an underwater camera (Schneider 1999). The identity of individuals performing the behavioral events was defined either through direct visual observations or from either photographs or videos. The marking rate (permanent nicks and notches on the dorsal fin) in this population is high (Williams et al. 1993; Currey et al. in press) which means that practically all individuals can be recognized from marks on their dorsal fins. Therefore practically all the population (excluding calves) was equally likely to be recognized in this way, minimizing sampling bias.

**Association patterns**

An association matrix of all individuals was obtained from the school membership samples, in which two individuals were associated if they were seen together in a school (Whitehead and Dufault 1999). We used a half-weight association index (Cairns and Schwager 1987) to estimate the proportion of time pairs of individuals spent associated. This index was used to account for small discrepancies in sampling effort between individuals (Lusseau et al. 2003). Only groups for which all individuals were identified were retained in the analysis. We obtained a weighted social network representation of the population's association pattern where edges between individuals represented their association index. This weighted social network provides more information about social relationships within the population than previously used binary social networks which categorized relationships as either existing or not (Lusseau and Newman 2004; Lusseau 2007a; Lusseau et al. 2008). However, not all network statistics used with binary networks can be readily applied to weighted networks (Lusseau et al. 2008). We therefore extended previous analyses using statistics for weighted networks which we describe below.

We defined clusters of individuals within the population, social units, using the modularity matrix clustering technique (Newman 2006b; Lusseau et al. 2008). The notion of modularity is based on defining a parsimonious division of the network which would allow maximizing the number (and weights) of edges within communities and minimizing the number, and weight, of edges between communities. A good cluster division provides many edges within clusters and few between (Newman and Girvan 2004). The modularity coefficient, Q, is the sum of associations for all dyads belonging to the same cluster minus its expected value if dyads associated at random, given the rate at which each individual in the dyad associated with all others in the population (their strength, Lusseau et al. 2008). This coefficient has the advantage of not disregarding the possibility that all individuals belong to only one cluster. Therefore the division which maximizes Q can be considered the "best" clustering of a network. The modularity matrix clustering algorithm uses, for each pair, the weight (association index in our case) between two vertices minus the expected weight if weights were randomly distributed (Newman 2006a). The eigenvector of the dominant eigenvalue of this modularity matrix provides a good division into two clusters (positive versus negative values on this vector). The technique is then used iteratively and the candidate community division is provided by the iteration that maximizes the modularity coefficient (Newman and Girvan 2004; Newman 2006b). Lusseau et al. (2008) provide more details about this method. From this analysis, we were able to determine social unit (community) membership for each individual and determine the degree of community mixing present in each school encountered. Each school was categorized with a mixing coefficient varying from 1, in cases where all individuals in the schools belonged to one unit, to $1/m$, $1/m$ of individuals belonged to one unit and the other school members belonged to others (where $m$ is the number of social units defined by the clustering analysis).

**Social network measures as 'knowledge' proxies**
In a previous study, some individuals were identified as social 'brokers' (Lusseau 2007a; Krause et al. 2009). Social 'brokers' had social relationships spread between clusters of individuals and were therefore likely to have a good knowledge of the activities of other clusters of individuals. Hence, they were relatively more likely to have knowledge about the food patches recently visited (and depleted) by other clusters of individuals than other individuals within their own cluster. Since patch depletion is crucial for patch quality in bottlenose dolphins, that means that social 'brokers' were likely to have personal information of the current quality of food patches that was not available to individuals that were not social 'brokers'. While the rate at which patches can 'recover' is unknown, we know that movement between patches is limited for several fish species (Lusseau and Wing 2006; Rodgers and Wing 2008). Therefore, replenishment of foraging patches is not only slower than it could be, but more importantly can also be predicted more accurately from depletion status alone.

It is important to note that specific socioecological conditions lead to the emergence of the usefulness of this information. Firstly, schools composed of members from both communities can be observed in all behavioural contexts. The ranging patterns of the communities are indistinguishable; hence, they do use the same foraging patches. Finally, members of a dolphin school, as in other echolocating mammals such as bats (Barclay 1982), can eavesdrop on each others to acquire information using the click trains produced by others (Dawson 1991; Götz et al. 2006). Therefore, the foraging experience of individual 1 can be passively transferred to individual 2 and hence individual 2 can be perceived as being more accurate in its decision-making process by individual 3 because it is more likely to have been around individuals with whom individual 3 does not interact.

To determine social 'brokers', we calculated the reach of individuals, a statistic that essentially highlights the same positions within the network as betweenness does in a binary network (Flack et al. 2006; Lusseau 2007a; Lusseau et al. 2008; Whitehead 2009). The reach of individual i is defined as the sum of the products of all association index pairs linking i and k through another individual j. The reach of an individual quantifies its indirect connectivity to others in the network, *i.e.* the number of individuals it can *reach* in the network at a given time (Newman 2003; Flack et al. 2006). It has also been used in defining opinion leaders in cooperation models on social networks (Eguiluz et al. 2005). In our case this statistic presents an understanding of the likelihood that an individual will have access to information about the whereabouts of others either actively or passively (because it spent more time with a greater diversity of individuals). There is therefore no assumption of active information transfer being measured by the reach statistic as information can as well be transferred passively (Seppänen et al. 2007). Randomization tests were used to estimate whether individuals that were observed performing SF and ULT were more likely to have a significantly higher reach than others given the amount of time we spent observing each individual in relation to the total amount of time we spent observing dolphin schools.

For each school we then subsequently identified the reach discrepancy in the school by measuring the maximum reach and the median reach of all individuals except the individual with the highest reach (median$_{OM}$ hereafter) (Conradt and Roper 2003). These school statistics are related to school size: the greater the school size, the more likely it is to include a high maximum reach, and subsequently the greater its median reach will be. However, this relationship is non-linear, reaching a plateau as school size increases, since reach values are upper bounded by the maximum reach. We therefore used an inverse function to relate maximum and median$_{OM}$ reach to school size (Eq. 1) and used the residual maximum and median$_{OM}$ reach for subsequent analyses.

$$reach = \alpha + \frac{\beta_1}{n} \quad \text{where } n \text{ is the school size} \tag{1}$$

This allowed us to fix the interaction terms between reach (maximum reach and median$_{OM}$) and school size to a relationship which could be expected by chance. Hence, the following models did not explore the influences of these variables on the occurrence of behavioral cues, but rather tested whether the predicted influences were observed, providing more predictive power to these models. It has to be noted that while reach can provide an understanding of the discrepancy of knowledge between individuals, it cannot measure absolute knowledge.

**Probability that unshared consensus decisions emerge**

We used logistic regressions to assess whether the probability of observing SF or ULT in a school depended on school parameters (Quinn and Keough 2002). Since SFs are performed by males and ULTs by females, we first contrasted the reach information of male members to compare SF schools and schools with no behavioral cues. We then compared the reach information of female members to compare ULT schools and schools with no behavioral cues. We also assessed whether school size, community mixing, and sex ratio influenced the probability that behavioral cues were performed.

For each regression a probability function was fitted using the logit link function:

$$p(y_i = 1) = \frac{1}{1+e^{-g(x)}} \qquad (2)$$

and $g(x) = \alpha + \beta_1 x_1 + ... + \beta_m x_m$ \qquad (3)

where $p(y_i = 1)$ is the probability that school *i* performed SF or ULT, α is a constant, and $\beta_1$ to $\beta_m$ are the coefficients of the independent variables $x_1$ to $x_m$. The odds ratio for each variable is given by Eq. 4.

$$odds_{x_i} = e^{\beta_i} \qquad (4)$$

A backward stepwise method was used to select the best fitting main effects ($x_1$ to $x_m$) with changes in the model's likelihood ratio being the criteria for variable selection. A forward stepwise method was then applied to confirm the selection from the backward method. Coefficients were estimated by maximum likelihood and the significance of each term was quantified by the change in the model's log-likelihood when that term was removed from the selected model.

Given the hypothesis, we would expect the residual maximum reach and the residual median$_{OM}$ reach to be retained in the regression models the former with a positive coefficient and the latter with a negative coefficient, highlighting greater reach discrepancies in ULT and SF schools.

**The dependence of decision-making process selection on contextual information**

Here, we develop a prediction about when bottlenose dolphins should reach unshared consensus decisions about travel timing in dependence of their absolute knowledge about optimal travel timing. The absolute knowledge of individuals can vary with the context within which the decision needs to be reached. The behavioral state of this bottlenose dolphin population is relatively spatially fixed. In addition to relying on reef-associated prey items (Lusseau and Wing 2006), they also tend to have socializing and resting hotspots which may be driven by the topography of the fiord and the likelihood to encounter predators (Lusseau and Higham 2004). Therefore all individuals are more certain about when it is optimal to stop traveling than about when it is optimal to start traveling because the former is more optimal when reaching a foraging/socializing/resting hotspot; it is spatially fixed. Once we account for the influence of contextual information in the Conradt-Roper decision-making model (Appendix 1), it appears that the more knowledgeable all individuals in the group are (even the individuals with low reach), the less steep a correlation would one expect between the discrepancy in reach within the school and the advantage of unshared consensus decision making, and, thus the likelihood of occurrence of unshared consensus decision making. Information about optimal travel initiation is likely to be low in low reach individuals because it is not linked to spatial contextual information in the same manner as traveling termination is. Hence, we predict a strong positive relationship between the occurrence of apparent unshared consensus SF decisions and the discrepancy in reach between males within a school. In contrast, information about optimal travel termination is likely to be generally high because it can be expected when dolphin schools encounter a behavioral hotspot. Therefore, we predict a less strong, but still positive relationship between the occurrence of apparent unshared consensus ULT decisions and the discrepancy in reach between females within a school.

# RESULTS

During the study period we obtained school membership information on 44 schools in which ULT was observed performed, 51 schools in which SF was observed, and 357 schools for which neither ULT nor SF was observed, but behavioral transitions were observed. The social network of the 53 adult individuals that were observed during this period was composed of 2 social units ($Q_{max}$=0.1, Figure 1) in agreement with previous analyses (Newman 2004; Rosvall and Bergstrom 2007; Monni and Li 2008) and observed behavioural variations in the population (Lusseau 2007b; Lusseau et al. 2008).

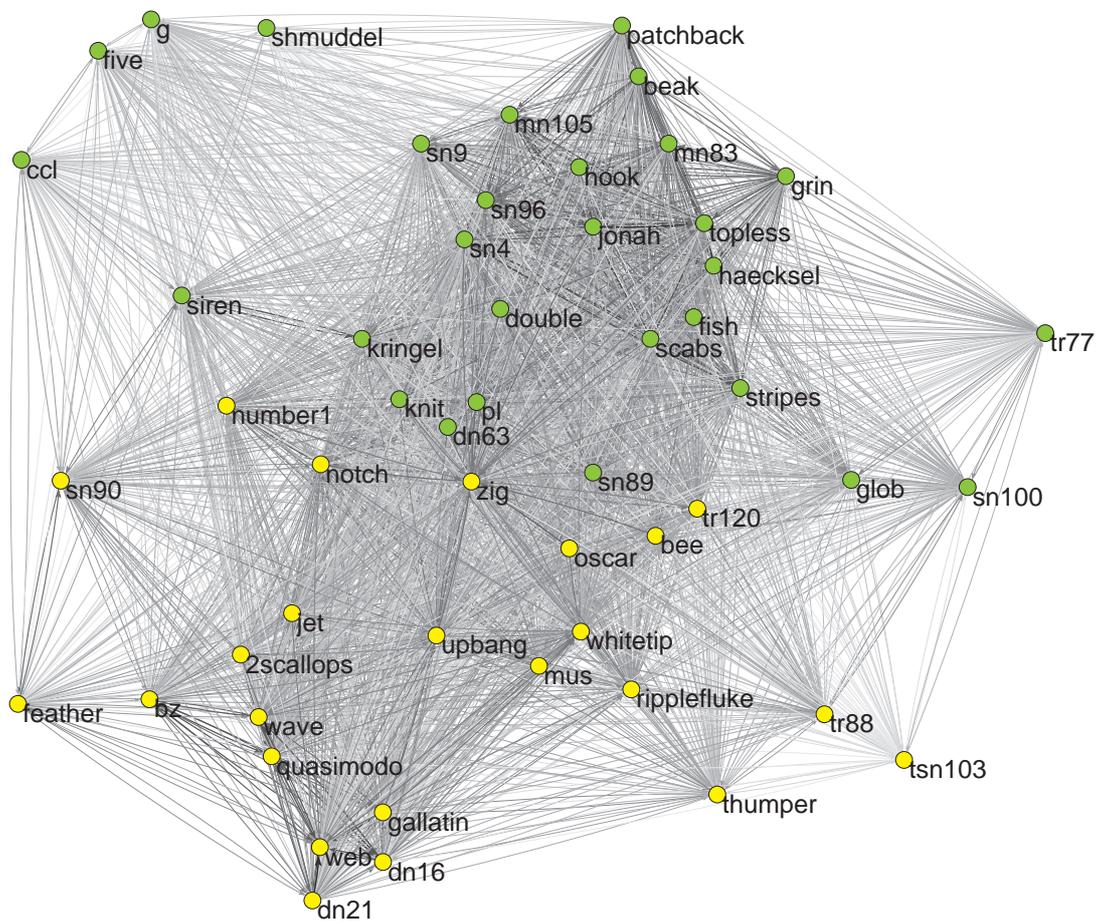

**Figure 1.** The weighted social network of the 53 individuals in this analysis: edges represent the association indices of the pairs and nodes are individuals, the darker the edge the greater the association index. Node color represents the community division obtained from the clustering analysis. Note that the network graph (a 'ridiculogram') does not provide much information about the structure of the network, which can then only be assessed using statistics.

Individual males that were identified performing SF had a significantly higher reach than other males (males: average reach$_{SF}$= 527±10.3 S.E., average reach$_{others}$= 439±15.7 S.E., p= 0.0027, effect size: Cohen's d = 1.34, 10000 randomizations) and individual females that were identified performing ULT also had significantly higher reach than other females (females: average reach$_{ULT}$= 433±27.0, average reach$_{others}$= 372±28.3, p= 0.0022, effect size: Cohen's d = 0.56, 10000 randomizations). In 8 out of the 10 instances in which the SF performer was identified, that individual either had the highest or the second highest reach in the school in which the behavior was observed. In 10% of cases where the ULT performer was identified she had the highest or second highest reach in the school. In all SF and ULT cases where the signaling individual was identified, the individual belonged to the community representing the majority of the school.

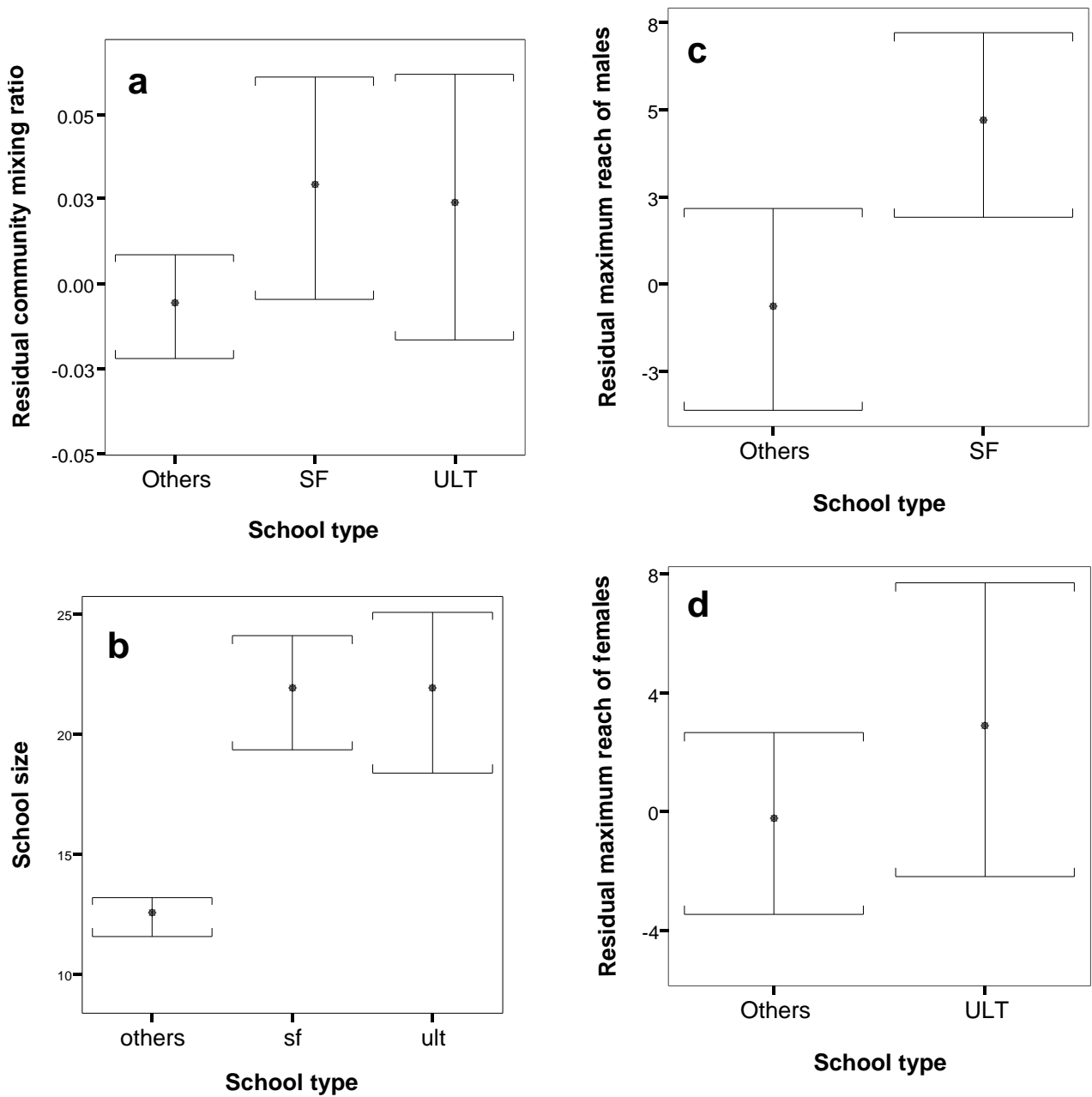

**Figure 2.** Variations in school characteristics depending on the performance of behavioral cues SF and ULT, all graphs represent the average with its 95% confidence interval: **a**. differences in school size; **b**. differences in community composition provided by the community mixing ratio, given the observed school size (residuals of a linear regression model between community ratio and school size, $F_{1,449}$=70.0, p<0.001, $R^2$=0.14) as community ratio is expected by chance to decrease as school size

increases; **c**. residual maximum reach of males in SF schools and in other schools (see methods); **d**. residual maximum reach of females in ULT schools and in other schools (see methods).

The reach of an individual did not depend on the number of times it was observed ($F_{1,52}$=1.8, p=0.18, $R^2$=0.03). Schools observed performing a behavioral cue tended to be larger than others ($F_{1,2}$= 47.9, p<0.001, Figure 2b).

The maximum reach and median$_{OM}$ reach of males in schools were non-linearly related to the number of males in the schools ($F_{1,369}$=145.6, p<0.001, $R^2$=0.28 and $F_{1,369}$=42.6, p<0.001, $R^2$=0.10 respectively). The maximum reach and median$_{OM}$ reach of females in schools were related to the number of females present in the schools ($F_{1,346}$=121.4, p<0.001, $R^2$=0.26 and $F_{1,346}$=46.3, p<0.001, $R^2$=0.12 respectively). We used the residuals from these models in the logistic regressions.

**Table 1.** Logistic regression model which describes the likelihood to observe SF being performed in a school. Reach statistics are the residual reach value given the inverse relationship between maximum and median reach and school size. Maximum and median$_{OM}$ values were calculated using only male members of the school. The significance of each variable is determined by the change in log-likelihood to the model if that term is removed.

| Variables | Odds ratio | 95% confidence interval | Change in log-likelihood | p |
|---|---|---|---|---|
| Constant | 0.001 | | | <0.001 |
| School size | 1.164 | 1.11-1.22 | 50.8 | <0.001 |
| Maximum reach | 1.036 | 1.01-1.07 | 8.3 | 0.004 |
| Community | 37.02 | 2.17-631.35 | 6.7 | 0.010 |
| Median$_{OM}$ reach | 0.99 | 0.98-0.999 | 4.8 | 0.028 |

The logistic regression model describing the probability that SF arises in a school retained all four independent variables (school size, community mixing, residual of maximum reach and residual of median$_{OM}$ reach, Table 1, Figures 2a-c). The model did not depart significantly from the data (maximum likelihood estimate of the model = 243.4; Hosmer & Lemeshow test $\chi^2_8$=7.4, p=0.49) and classified 85.7% of schools correctly. In contrast, the logistic regression model describing the probability that ULT arises in a school only retained school size and community mixing as terms and the latter did not provide a significant explanation of the data variance (Table 2, Figures 2a, b, d). That model did not depart significantly from the data (maximum likelihood estimate of the model = 225.6; Hosmer & Lemeshow test $\chi^2_8$=7.6, p=0.47) and classified 88.5% of schools correctly.

**Table 2.** Logistic regression model which describes the likelihood to observe ULT being performed in a school. The significance of each variable is determined by the change in log-likelihood to the model if that term is removed.

| Variables | Odds ratio | 95% confidence interval | Change in log-likelihood | p |
|---|---|---|---|---|
| Constant | 0.003 | | | <0.001 |
| School size | 1.122 | 1.07-1.17 | 30.1 | <0.001 |
| Community | 9.71 | 0.64-147.64 | 2.8 | 0.096 |

## DISCUSSION

One limitation of this study is that we have no formal demonstration that SF and ULT cause the school to shift activity. While field observations show a distinct change in activity right after the performance of these behavioral cues, causal relationships could only be inferred from experimental studies such as playback experiments. This lack of causal relationship demonstration is shared by most studies of decision-making processes in animals (Conradt and Roper 2005). It is highly possible that while these behaviors are primarily acoustic cues, they may also be perceived visually or in more complex ways, such as is the case with aerial behaviors in spinner dolphins (*Stenella longirostris*) (Norris 1994). It would therefore be difficult to set up experiments to demonstrate a causal relationship; a limitation shared with other examples of animal consensus-reaching signals. However, the energetic cost of these signals, especially side flops, is such that they would be an honest portrayal of the motivational state of the performing individual (Bradbury and Vehrencamp 2000). In addition, since these two events are not performed in any other contexts than activity state shifts, their performance is closely linked to shift in the school's behavioral state (Lusseau 2006). It is possible that these cues are only used to reinforce another, acoustic, signal to increase the robustness of its transfer in situations where the primary signal may be occluded (Ay et al. 2007). This situation could easily arise in larger schools, where low frequency sounds produced by SF and ULT could reinforce high frequency vocalizations to ensure that the signal reaches all individuals in the school. School size was one of the best predictor explaining the emergence of SF and ULT in regression models (Tables 1 and 2) which supports this hypothesis. However, acoustic studies did not find any vocalizations more likely to be performed during state shifts (Boisseau 2004; Boisseau 2005).

Individuals observed performing SF and ULT were more likely to have a high reach value than others, which confirmed results from previous studies (Lusseau 2007a). However, while SF performers were highly likely to have the highest reach in the school in which they performed SFs, it was not the case for ULT performers. Indeed the reach of individuals in a school seemed to play little role in the observation of ULTs. These were more likely to be observed in larger schools composed mostly of individuals coming from the same social unit. While this was also true for SF occurrence, SF schools had in addition a greater reach discrepancy between the individual male with the highest reach and other males in the school. This latter result fits the expected requirements for the emergence of unshared consensus decisions in Conradt & Roper's general model (2003), and in our specific model (Appendix 1).

There may be several reasons for the discrepancies in results between ULT and SF performances. Firstly, it is possible that reach better tracks the 'knowledge content' of males than the one of the female. A long time period is required to assemble enough school membership samples to obtain precise reach estimations for each individual. In the case of this study reach value is approximated over a two-year period. Whether this value will be useful to approximate the information content of an individual will depend on how temporally consistent this content is. In instances where the relevance of the information gathered expires over a short time scale, the reach statistic may be able to point out individuals more likely to hold relevant information (the brokers), but which of those individuals possess relevant information at time *t* could not be ascertained by the reach statistic. SFs are performed to start travel by males and ULTs by females to stop traveling. We previously hypothesized that this pattern may be linked to the difference in the cost of transport between the two sexes and the need to maintain mixed-sex schools (Conradt and Roper 2000; Lusseau et al. 2003; Lusseau 2007a). Following this reasoning we would not expect to see any differences in the way reach tracks the information content of males and females, especially since signaling females still have significantly higher reach than others overall.

Secondly, both SF and ULT were more likely to be performed in schools composed mainly of individuals from the same social unit. An honest signal, performed by an informed individual, can lead to unshared decisions in large groups (Couzin et al. 2005). In our cases the signaling individual would have vested interests in honestly displaying to its community co-members, and in large groups a proportionally smaller number of individuals can sway the decision of the whole group (Couzin et al. 2005). ULT were also performed in contexts where individuals were generally more likely to be knowledgeable about optimality of activity shifts than in the case of SF. According to our model predictions (Appendix 1), if knowledge about optimal timing is generally high, the expected increased in emergence of a unshared consensus decision with a heightened reach discrepancy between the decision-maker and others in the school is shallow, and could therefore, have easily been masked by noise in our data.

In conclusion, it is possible for unshared consensus decisions to be reached when, as predicted by Conradt and Roper's model (2003), the individual making the decision has a greater knowledge about the benefits of shifting activities than all other members of a group. However, this may not be the only process under which unshared consensus decisions may be reached and other mechanisms can be at play. We have some indications here that unshared consensus can emerge when the expectation of a decision is high for all individuals, leading to a smaller cost of selecting despotism over democracy.

## Acknowledgements

DL was supported by a Killam Postdoctoral Fellowship provided by the Killam trusts. LC is supported by a Royal Society University Research Fellowship. We would like to thank Hal Whitehead for numerous fruitful discussions and suggestions, Shane Gero for suggestions on earlier drafts and three anonymous reviewers. Data collection and compilation was funded by the New Zealand Whale and Dolphin Trust, the New Zealand Department of Conservation, Real Journeys Ltd, and the University of Otago (Departments of Zoology and Marine Sciences and Bridging Grant scheme). We would also like to thank Susan M. Lusseau, Oliver J. Boisseau, Liz Slooten, and Steve Dawson for their numerous contributions to this research.

**Appendix 1. Contextual knowledge and its influence on the emergence of unshared consensus decisions**

The following model is based on the Condorcet jury theorem. Its purpose is to show that if the level of information is generally high within the group, any increase in information of the group member with the highest reach (in relative and in absolute terms) adds relatively little advantage to an unshared decision. On the other hand, if the level of information is generally low within the group, already a small difference in information between the highest reach member and other group members might make an unshared decision more profitable than a shared decision. This could explain the different observations in decisions about the initiation and about the termination of traveling bouts, if these two types of decisions are accompanied by different general levels of relevant information within the group.

We assume that the animal with the highest reach in a school has the probability $p_r$ to get the decision about timing of traveling right (i.e., to time it optimally given their energy budget), and the probability 1- $p_r$ to get it wrong (i.e., time the traveling not optimal). Further, we assume that all other group members, which have lower reach, have a lower probability $p_s$ ($p_s < p_r$) to get the decision right, if they made the decision individually. Thus, the median probability to get the decision right of all individuals is $p_s$ (assuming group size is larger than two). This median probability $p_s$ is, thus, a measure of 'the general level of information within the group'.

We rewrite $p_r$ as:

$p_r = p_s + \lambda \times (1 - p_s)$, with $0 < \lambda \leq 1$ (since $p_r > p_s$, and $p_r$ is bounded at one) (1)

Hence,

$p_r - p_s = \lambda \times (1 - p_s)$ (2)

$\lambda$ is a measure of the difference in information between the most informed group member and the other members.

The probability that a school would make a right decision if it followed the SF (or ULT, respectively) signal of the animal with highest reach, would be $p_r$ (unshared consensus decision). The probability that a school would make a right decision if it decided democratically (assuming for reasons of simplicity that school size $n$ is uneven), would be:

$$p_r \cdot \sum_{i=\frac{n-1}{2}}^{n-1} \binom{n-1}{i} \cdot p_s^{i} \cdot (1-p_s)^{n-1-i} + (1-p_r) \cdot \sum_{i=\frac{n+1}{2}}^{n-1} \binom{n-1}{i} \cdot p_s^{i} \cdot (1-p_s)^{n-1-i} \qquad (3)$$

Thus, a school would make a better decision by following the most knowledgeable individual, if:

$$p_r > p_r \cdot \sum_{i=\frac{n-1}{2}}^{n-1} \binom{n-1}{i} \cdot p_s^{i} \cdot (1-p_s)^{n-1-i} + (1-p_r) \cdot \sum_{i=\frac{n+1}{2}}^{n-1} \binom{n-1}{i} \cdot p_s^{i} \cdot (1-p_s)^{n-1-i} \qquad (4)$$

and the difference between an unshared and a shared consensus decision ($diff_{desp\text{-}dem}$) in terms of probability to make the right decision would be:

$$diff_{desp\text{-}dem} = p_r - p_r \cdot \sum_{i=\frac{n-1}{2}}^{n-1} \binom{n-1}{i} \cdot p_s^{i} \cdot (1-p_s)^{n-1-i} - (1-p_r) \cdot \sum_{i=\frac{n+1}{2}}^{n-1} \binom{n-1}{i} \cdot p_s^{i} \cdot (1-p_s)^{n-1-i} \quad (5)$$

$$= p_r - p_r \cdot \sum_{i=\frac{n-1}{2}}^{\frac{n-1}{2}} \binom{n-1}{i} \cdot p_s^{i} \cdot (1-p_s)^{n-1-i} + p_r \cdot \sum_{i=\frac{n-1}{2}+1}^{n-1} \binom{n-1}{i} \cdot p_s^{i} \cdot (1-p_s)^{n-1-i}$$

$$- (1-p_r) \cdot \sum_{i=\frac{n+1}{2}}^{n-1} \binom{n-1}{i} \cdot p_s^{i} \cdot (1-p_s)^{n-1-i} \quad (6)$$

$$= p_r \cdot \left[ 1 - \frac{(n-1)!}{\left(\frac{n-1}{2}!\right)^2} \cdot p_s^{\frac{n-1}{2}} \cdot (1-p_s)^{\frac{n-1}{2}} \right] - \sum_{i=\frac{n+1}{2}}^{n-1} \binom{n-1}{i} \cdot p_s^{i} \cdot (1-p_s)^{n-1-i} \quad (7)$$

$$= (p_s + \lambda \cdot (1-p_s)) \cdot \left[ 1 - \frac{(n-1)!}{\left(\frac{n-1}{2}!\right)^2} \cdot p_s^{\frac{n-1}{2}} \cdot (1-p_s)^{\frac{n-1}{2}} \right] - \sum_{i=\frac{n+1}{2}}^{n-1} \binom{n-1}{i} \cdot p_s^{i} \cdot (1-p_s)^{n-1-i} \quad (8)$$

Since $diff_{desp\text{-}dem}$ is the difference between an unshared and a shared decision with respect to the probability of getting the decision right, it is a measure of the relative advantage of an unshared over a shared decision.

From equation (8), it follows that $diff_{desp\text{-}dem}$ is correlated with the difference in information between the most informed group member and the other members (λ) as follows:

$$\frac{\partial diff_{desp-dem}}{\partial \lambda} = (1-p_s) \cdot \left[ 1 - \frac{(n-1)!}{\left(\frac{n-1}{2}!\right)^2} \cdot p_s^{\frac{n-1}{2}} \cdot (1-p_s)^{\frac{n-1}{2}} \right] \quad (9)$$

That is, the advantage of an unshared versus a shared decision increases with the information discrepancy λ with a slope given by Eq. 9. The size of this slope is always positive, but decreases with $p_s$ (Figure 3). That means, the larger the general level knowledge of all group members within the group (i.e., $p_s$), the less advantage (i.e., $diff_{desp\text{-}dem}$) does an increase in superior information of the member with the highest reach (i.e., λ) convey in an unshared decision; and *vice versa*.

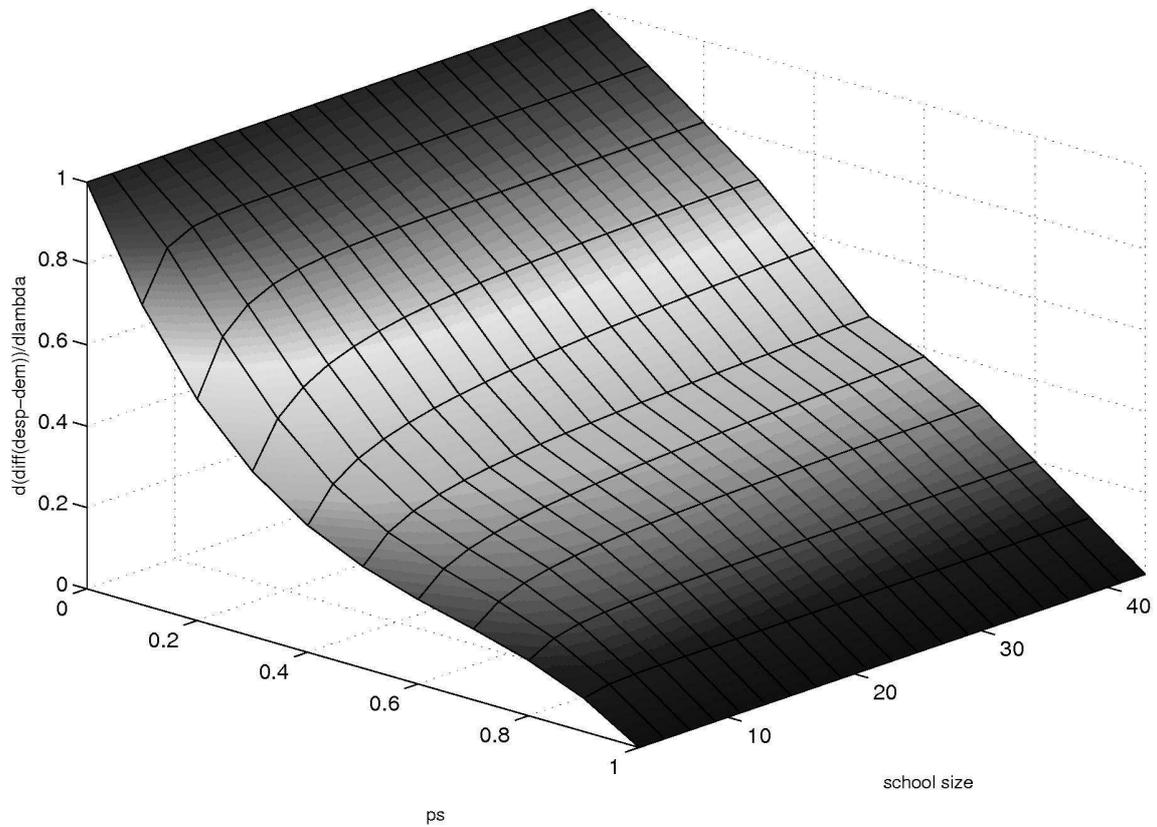

**Figure 3.** The relationship between the median knowledge of members of the schools ($p_s$, i.e. excluding the individual with the maximum knowledge), school size, and $\dfrac{\partial \mathit{diff}_{desp-dem}}{\partial \lambda}$ which is the difference in correctness to take an unshared consensus decision as opposed to a shared consensus one ($\mathit{diff}_{desp-dem}$) given the discrepancy in knowledge between the most knowledgeable and others in the schools ($\lambda$). See Appendix 1 for derivation.